\documentclass[showpacs,preprintnumbers,amsmath,amssymb,pre,twocolumn]{revtex4-1}
\usepackage{hyperref}
\usepackage{subfigure}
\usepackage{graphicx}
\usepackage{dcolumn}
\usepackage{bm}
\usepackage{hyperref}
\usepackage{multirow}
\usepackage{url}
\usepackage[latin1]{inputenc}
\usepackage{amssymb}
\usepackage{array}
\usepackage{hhline}
\usepackage{longtable}

\begin{document}


\title{A universal assortativity measure for network analysis}

\author{Guo-Qing Zhang$^{1}$}
\email{gqzhang@ict.ac.cn}
\author{Su-Qi Cheng$^{1,3}$}
\email{chengsuqi@ict.ac.cn} 
\author{Guo-Qiang Zhang$^{1,2}$}
\email{guoqiang@ict.ac.cn}

\affiliation{$^{1}$Institute of Computing Technology, Chinese Academy of Sciences}
\affiliation{$^{2}$School of Computer Science and Technology,
Nanjing Normal University}
\affiliation{$^{3}$Graduate University of
Chinese Academy of Sciences}



\begin{abstract}
Characterizing the connectivity tendency of a network is a
fundamental problem in network science. The traditional and
well-known assortativity coefficient is calculated on a per-network
basis, which is of little use to partial connection tendency of a network
. This paper proposes a universal assortativity coefficient
(UAC), which is based on the unambiguous definition of each
individual edge's contribution to the global assortativity
coefficient (GAC). It is able to reveal the connection tendency of
microscopic, mesoscopic, macroscopic structures and any given part of a network. Applying UAC to
real world networks, we find that, contrary to the popular expectation, most
networks (notably the AS-level Internet topology) have markedly more
assortative edges/nodes than dissortaive ones despite their global
dissortativity. Consequently, networks can be categorized along two
dimensions--single global assortativity and local assortativity
statistics. Detailed anatomy of the AS-level Internet topology
further illustrates how UAC can be used to decipher the hidden
patterns of connection tendencies on different scales.


\end{abstract}

\maketitle


\section{Introduction}

Network has become a useful and proliferative tool in a wide
spectrum of research areas, ranging from traditional communication
and transportation networks to more recently emerging networks as
complex as online social networks and brain
networks~\cite{barabasi1999,faloutsos1999,dorogovtsev2002,albert2002,newman2003,mislove2007,bullmore2009,zhang2008,zhang2011,spring2002,newman2001a,newman2001b,dorogovtsev2004,zhang2007,yan2006}.
Assortativity coefficient is a basic metric that characterizes the
connectivity tendency of a network, i.e., globally, whether nodes of
similar(or dissimilar) degrees are more likely to be
connected~\cite{newman2002}. However, this metric is a macroscopical
property, which becomes useless when microscale or mesoscale level
analysis is required. In other words, one can not tell the exact
intra-group or inter-group connection tendencies from the
per-network assortativity coefficient.

\begin{figure}
\centering
\includegraphics[scale=0.4]{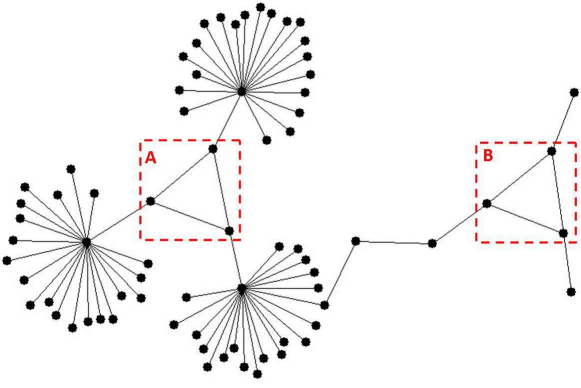}
\caption{An illustrative example showing how local connectivity
pattern differentiate two isomorphic components(ie. A and B). Different local connectivity patterns always have different effects on dynamics, such as disease spreading.
\label{example}}
\end{figure}

Experimental studies have shown that various forms of groups
are hidden in real networks. These groups can take the form of
community, motif, clique, etc~\cite{girvan2002,newman2004,milo2002}.
Multi-scale, especially mesoscale analysis is very important to
understand the roles and dynamics of these
groups~\cite{arenas2008,almendral2011}. However, previous studies
typically focus on the uncovering of these groups within a network,
and treat isomorphic modular components to be identical.
In other words, the component is solely studied as a subgraph
extracted out of the whole network, totally neglecting the links
connecting this subgraph to other parts of the graph. Obviously, this traditional method
inevitably fails to capture the functional difference between
isomorphic modular components. Indeed, functional roles
or dynamics of a group can only be comprehensively
understood when it is put in the global context. An important
distinguishable property is whether the group under consideration is
assortatively mixed or dissortatively mixed within its local
surroundings, which can have quite different influence on
the dynamics, e.g., information diffusion/disease
spreading~\cite{kitsak2010}, resilience against
attacks~\cite{albert2000}. 
 Fig.~\ref{example} gives an illustrative example. In this
figure, two triangles $A$ and $B$ are located in different
surroundings. Triangle $A$ is surrounded by high-degree nodes, i.e.,
dissortatively mixed with the outside world,  whereas triangle $B$
is surrounded by low-degree nodes, i.e., assortatively mixed with
the outside world. This causes $A$ and $B$ to behave quite
differently in the process of information or disease diffusion. In
this simple example, suppose SIR model is used to model a disease
spreading process and the infectious probability $p$ is set to 0.5.
If $A$ serves as the source of the spreading process, the expected
number of infected nodes accounts for about 23\% of all the nodes,
in contrast, if $B$ serves as the source, then only less than 4\% of
the nodes are expected to be infected. This drastic discrepancy
apparently comes from the difference in the connectivity tendency
between the two triangles and their respective outside worlds.

Hence, in order to exactly analyze and explore network structure, which
is beneficial to better understand the dynamics of complex
systems, it is of critical significance to perform intra-group or
inter-group connection tendency measurement in the global context.
In this paper, we propose a universal assortativity coefficient that
is based on the unambiguous definition of each individual edge's
contribution to the global assortativity coefficient. This metric
allows assortativity analysis on any part of a network and
reveals some hidden network connectivity patterns.

\section{Universal Assortativity Coefficient}
In order to measure the assortativity of the network on different
scales, we proposed a uniform metric called universal assortativity
coefficient that measures the assortativity of any subset of
connections. Simply put, it is the summation of each individual
edge's contribution to the global assortativity coefficient. Hence,
we begin with our definition of each individual edge's contribution
to the global assortativity coefficient.

Before the formal definition, it is necessary to review some related
concepts discussed by Newman ~\cite{newman2002}. For simplicity, all
the concepts we discuss are based on undirected networks. With minor
or moderate adjustments, these concepts can also be applied to
directed networks. Degree distribution $p(k)$ refers to the
probability that a randomly chosen node is of degree $k$. The
remaining degree distribution $q(k)$ refers to the probability that
following a randomly chosen edge, the remaining degree of the
reached node is $k$. Here, the remaining degree is the number of
edges leaving this node other than the one we arrived along. This
number is one less than the total degree of this node. The
normalized distribution $q(k)$ of the remaining degree is:
\begin{equation}
q(k)=\frac{(k+1)p(k+1)}{\Sigma jp_j}
\end{equation}

Joint probability distribution of the remaining degrees of two
endpoints at either end of a randomly chosen edge $e_{ij}$ is the
probability that the remaining degrees of two endpoints of a
randomly chosen edge are $i$ and $j$.

Following these definitions, the assortativity coefficient $r$ is
defined as:
\begin{equation}
r=\frac{1}{\sigma_q^2}[\Sigma_{jk}jk(e_{jk}-q(j)q(k))]
\end{equation}
where $\sigma_q$ is the standard deviation of the remaining degree
distribution $q(k)$.

For uncorrelated network, $r$=0; when the network is assortatively
mixed, i.e., nodes of similar degrees are more likely to get
connected, $r$ is positive; when the network is dissortatively
mixed, i.e., nodes of dissimilar degrees tend to connect to each
other, $r$ is negative.

Now considering each individual edge's contribution to the network
assortativity coefficient $r$.  Denote $U_q=\Sigma_j{jq(j)}$ to be
the expected value of remaining degree, then $r$ can be rewritten
as:
\begin{equation*}
\begin{split}
r & = \frac{1}{\sigma_q^2}[\Sigma_{jk}jk(e_{jk}-q(j)q(k))]\\
  & = \frac{\sigma_{jk}{jk}e_{jk}-U_q^2}{\sigma_q^2}\\
  & = \frac{\sigma_{jk}jke_{jk}-\Sigma_jjq(j)-\Sigma_kkq(k)+U_q^2}{\sigma_q^2}\\
  & = \frac{\sigma_{jk}jke_{jk}-\Sigma_jj\Sigma_kke_{jk}-\Sigma_kk\Sigma_jje_{jk}+U_q^2}{\sigma_q^2}\\
  & = \frac{\Sigma_{jk}jke_{jk}-\Sigma_{j}\Sigma_{k}(j+k)e_{jk}+U_q^2}{\sigma_q^2}\\
  & = \frac{\Sigma_{jk}{(jk-(j+k)U_q+U_q^2)e_{jk}}}{\sigma_q^2}\\
  & = \frac{\Sigma_{jk}{(j-U_q)(k-U_q)e_{jk}}}{\sigma_q^2}\\
  & = \frac{E(J-U_q)(K-U_q)}{\sigma_q^2}
\end{split}
\end{equation*}
where $J$ and $K$ are variables of the remaining degree, which have
the same expected value $U_q$. Following the above equation, we see
that each edge's contribution to $r$ is :
\begin{equation}
\rho_e=\frac{(j-U_q)(k-U_q)}{M\sigma_q^2}
\end{equation}
where $M$ is the number of edges, and $j, k$ are the remaining
degrees of the two endpoints of edge $e$. It is easy to see that
$r=\sum_{i=1}^{M}\rho_e$.

When the network is completely homogeneous, i.e., all nodes have the
same degree, then $\sigma_q=0$. In this case $\rho_e$ becomes
undefinable. Since in this case, each edge has the same contribution
to $r$, we define $\rho_e$ to be $\frac{1}{M}$.

If $\rho_e>0$, then $e$ is called an assortative edge; otherwise if
$\rho_e<0$, it is called a dissortative edge. In this
definition, if both the endpoints' remaining degrees are greater(or
less) than the global expected remaining degree $U_q$, then the edge
is assortative, and the more the two endpoints' remaining degrees
deviate from $U_q$, more assortative the edge is. Otherwise, the
edge is dissortative. In other words, the edge assortativeness is a
scaled difference between the two endpoints' remaining degrees and
the global expected remaining degree.The absolute
value of the contribution $|\rho_e|$ is termed as the
assortative/dissortative strength of the corresponding edge. We
define $S_{ae}$ to be the average strength of assortative edges, and
$S_{de}$ to be the average strength of dissortative edges. The ratio
of assortative edges is denoted by $P(\rho_e>0)$.

Finally, the universal assortativity coefficient for a targeted edge
set $E_{target}$ is defined as:
\begin{equation}
\rho=\sum_{e\in E_{target}}\rho_e=\sum_{e\in
E_{target}}\frac{(j-U_q)(k-U_q)}{M\sigma_q^2}
\end{equation}

 Based on this metric, it is easy to measure the assortativity on
different scales. For example, to measure the connectivity tendency
of a single node, denoted as $\rho_v$, simply set $E_{target}$ to be
the edges emanating from the node. If $\rho_v>0$, then we call $v$
to be an assortative node, otherwise if $\rho_v<0$, we call $v$ to
be a dissortative node.
To measure the connectivity tendency within a group, $E_{target}$ is
set to the edges within this group. If we set $E_{target}$ to be the
whole edge set $E$, then we arrive at the Newman's global
assortativity coefficient~\cite{newman2002}. In order to measure the
connectivity tendency between groups, simply set $E_{target}$ to be
the edges between the groups. In this sense, this metric can be used
to measure connection tendencies on different scales, thus, it
deserves the name uniform assortativity coefficient (UAC).

Back to our example in Fig.~\ref{example}, the global
assortative coefficient $\rho$ is -0.804, indicating strong
dissortativity. However, this global knowledge is of little use to
understand the functional roles of local components, such as $A$ and
$B$. Based on $UAC$, we can quantitatively measure the connection
tendency between $A$ and the remaining graph, as well as between $B$
and the remaining graph. It turns out that the inter-group
assortative coefficient between $A$ and the remaining graph is
-0.033, whereas the inter-group assortative coefficient between $B$
and the remaining graph is 0.025. As a consequence, although $A$ and
$B$ are isomorphic when they are extracted out of the graph, their
different connectivity tendencies to the other part of the graph
result in drastic discrepancy in the disease spreading process. This example
clearly tells us the significance of partial connection tendency for network analysis.


\section{Real network analysis}
\begin{table*}
\centering \caption{Connection tendencies for different categories
of networks.\footnote{The result of ER network is an average over 10
times. We treat Soc-Epinionsl, celegansneural and foodweb\_Florida,
originally directed networks, as undirected networks by treating
each directed edge as an undirected one and eliminating duplicated
edges.} \label{table-1}}

\begin{tabular}{|c|c|c|c|c|c|c|c|}

 \hline \hline

Category          & Name        &  $r$    & $ P(\rho_e>0) $  & $S_{ae}$          &  $  S_{de}  $ & $P(\rho_v>0)$ \\
\hline

\multirow{2}{*}{Technical Network} & AS-2011-6~\cite{asgraph} & -0.184  & 60.4\%             & $1.96\times10^{-6}$ &  $8.71\times10^{-6}$ & 58.3\%\\
                  & Router~\cite{spring2002}      & -0.138  & 51.3\%            & $5.80\times10^{-5}$ &  $1.07\times10^{-4}$ & 57.5\%\\
                  & USAir~\cite{usair,batagelj2006}                       & -0.208  & 42.1\%            & $2.76\times10^{-4}$ & $3.70\times10^{-4} $ & 37.3\% \\

\hline

\multirow{3}{*}{Biological Network}   & PPI~\cite{ppi}     & -0.102  & 50.5\%            & $6.92\times10^{-5}$ &  $1.02\times10^{-4}$ & 52.6\%\\
                                   & celegansneural~\cite{watts1998,white1986}     & -0.163  &
56.7\%  & $1.11\times10^{-4}$ & $3.21\times10^{-4}$ & 42.1\% \\
                                   & foodweb\_Florida~\cite{foodweb,batagelj2006}    & -0.112  &
49.6\%  &$1.90\times10^{-4}$ & $2.94\times10^{-4}$ & 34.4\% \\

 \hline

\multirow{3}{*}{Social Network}    & SCN~\cite{newman2001a,newman2001b}        & 0.161   & 61.4\%            & $1.33\times10^{-5}$ &  $1.07\times10^{-5}$ & 69.7\%\\
                                   & CA-HepTh~\cite{leskovec2007}    & 0.268   &
63.1\%
&$2.78\times10^{-5}$ & $1.96\times10^{-5}$ & 72.8\% \\
                                   & CA-GrOc~\cite{leskovec2007}     & 0.659   &
80.6\%
                  & $6.32\times10^{-5}$ & $2.78\times10^{-5}$ & 88.9\% \\
\hline

\multirow{2}{*}{Online Social Network}        & soc-Epinions1~\cite{richardson2003}& -0.041 & 58.8\%            & $5.81\times10^{-7}$ &  $1.07\times10^{-6}$ & 71.6\%\\
                                              & Email-Enron~\cite{klimmt2004,leskovec2009}  & -0.111  & 58.9\%           & $1.21\times10^{-6}$ &  $3.19\times10^{-6}$ & 51.5\%\\

\hline

Synthesized Network & ER~\cite{erodos1959}         & -0.001 & 50.7\%            & $1.24\times10^{-5}$ &  $1.27\times10^{-5}$ & 50.4\% \\
\hline
\end{tabular}
\end{table*}

We apply the UAC analysis to various real-world networks. Table.
~\ref{table-1} reports $r$, $P(\rho_e>0)$, $S_{ae}$, $S_{de}$ and
$P(\rho_v>0)$ for different kinds of networks. These networks can be
roughly categorized as five kinds: technical networks, biological
networks, social networks, online social networks, and synthesized
networks.

From this table, we see:
\begin{enumerate}
\item For a majority of real networks considered in this paper, e.g., AS, Router, Email-Enron,  despite their impressive global dissortativity, we
surprisingly find that the number of assortative edges/nodes exceeds
dissortative edges/nodes. Whereas for the synthesized ER network,
the number of assortative edges almost equals that of dissortative
edges, and their average strengths are indistinguishable as well.
Hence, the network as a whole has no mixing pattern.
\item The global network assortativity is determined by both the
ratio of assortative edges and the strength of these edges. For
instance, in SCN, both the ratio of assortative edges and the
average strength of assortative edges are greater than dissortative
edges, hence it exhibits strong assortativeness as a whole. In
comparison, though the number of assortative edges in the AS network
also exceeds dissortative ones, the average strength of assortative
edges is much weaker than dissortative ones. Hence, the
dissortativity of this network comes from the relatively stronger
strength of smaller number of dissortative edges. This is true for
quite a number of other dissortative networks.
\item Here we reconfirm the fact that online social networks are dissortatively
mixed, whereas real-world social networks are assortatively
mixed~\cite{hu2009}. We observe that the ratio of assortative edges
in online social networks are comparatively lower than that of
real-world social networks, although the total number of assortative
edges still exceeds dissortative ones. However, the average strength
of dissortative edges is greater than that of assortative edges in
online social networks, in contrast, in real-world social networks,
the situation is just the opposite. This reflects the fact that
online social networks can to some extent eliminate social barrier
between people of different social positions, making it is much
easier for people at the bottom of society to setup links to people
at the top of society.
\item According to global assortativity and local edge
assortativity statistics, networks can be categorized to four kinds:
globally assortative with leading number of assortative edges,
globally assortative but with leading number of dissortative edges,
globally dissortative with leading number of dissortative edges,
globally dissortative but with leading number of assortative edges.
Table~\ref{network_categorization} categorizes the networks along
the two dimensions. Yet, it still remains an open question whether
there is a real network that exhibits global assortativety but
primarily consists of dissortative edges.

\end{enumerate}

\begin{table}
\centering \caption{Categorization of networks by global
assortativity and edge assortativity
statistics.\label{network_categorization}}
\begin{tabular}{|c|c|c|}
\hline\hline
   & $r>0$  & $r<0$ \\
\hline

\multirow{3}{*}{$P(\rho_e>0)>50\%$}  & SCN             & AS, Router  \\
                                     & CA-HepTh        & soc-Epinionsl \\

                                     & CA-GrOc         & Email-Enron \\
 \hline
\multirow{2}{*}{$P(\rho_e>0)<50\%$}  & \multirow{2}{*}{-}  &  USAir\\
                                     &   & foodweb\_Florida \\
\hline
\end{tabular}
\end{table}

\begin{figure}
\centering
\includegraphics[width=8cm]{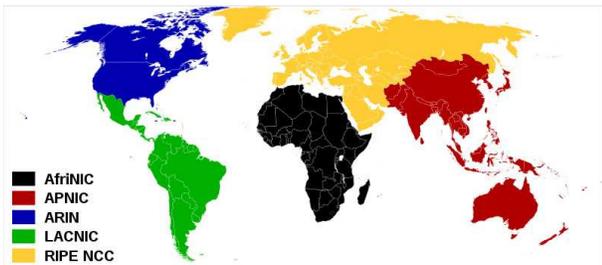}
\caption{(Color Online) Map of regional Internet registries.
\label{map}}
\end{figure}

In the following, we use the AS-level Internet topology as an
example to illustrate how the universal assortativity coefficient
can be used to calculate the connectivity tendency of intra-group or
inter-group connections. In the AS-level topology, a natural group
partition of clear and explicit meaning is to partition the ASes
according to their geographical regions. Today, five regional
Internet registries (RIR) are managing the allocation and
registration of Internet number resources (including AS numbers)
within a particular region of the world. The five RIRs are: AfriNIC
for Africa, ARIN for the United States, Canada, several parts of the
Caribbean region and Antarctica, APNIC for Asia, Australia, New
Zealand, and neighboring countries, LACNIC for Latin America and
parts of the Caribbean region, and RIPENCC for Europe, the Middle
East and Central Asia (see Fig.~\ref{map} for a graphical
representation of the five RIRs' responsible regions). This gives us
a coarse partition of the ASes according to the five regions. A more
fine-grained partition is to further divide each region according to
countries and regions. Hence, we have a two-level partitioning. The
first-level groups consist ASes adhering to the same regional
Internet registries, and the second-level groups consist ASes
belonging to the same country and region, following the ISO 3166-1
standard.

\begin{table*}
\centering \caption{Intra-RIR and inter-RIR assortativity
coefficients. \label{inter-RIR}}
\begin{tabular}{|c|c|ccccc|}
\hline\hline
   size        &  &     AfriNIC       &      APNIC       &        LACNIC      &       RIPENCC        &       ARIN      \\
\hline

380 & AfriNIC
 &  $9.35\times10^{-4}$  &  $3.86\times10^{-5}$  &  $1.13\times10^{-5}$  &  $1.24\times10^{-4}$  &  -0.002
\\
3711 & APNIC     & $3.86\times10^{-5}$ &     0.014 &
$1.48\times10^{-4}$& $-2.93\times10^{-4}$ & -0.01 \\
1209  & LACNIC    & $1.13\times10^{-5}$ &$1.48\times10^{-4}$ & 0.004
&
$-4.61\times10^{-4}$ & -0.007 \\
13401 & RIPENCC &  $1.24\times10^{-4}$  & $-2.93\times10^{-4}$ &
$-4.61\times10^{-4}$       & 0.021
&  -0.083  \\
11172 & ARIN      &    -0.002   &   -0.01  &  -0.007 & -0.083
 & -0.121 \\
\hline
\end{tabular}
\end{table*}

\begin{figure*}
\centering
\includegraphics[scale=0.6]{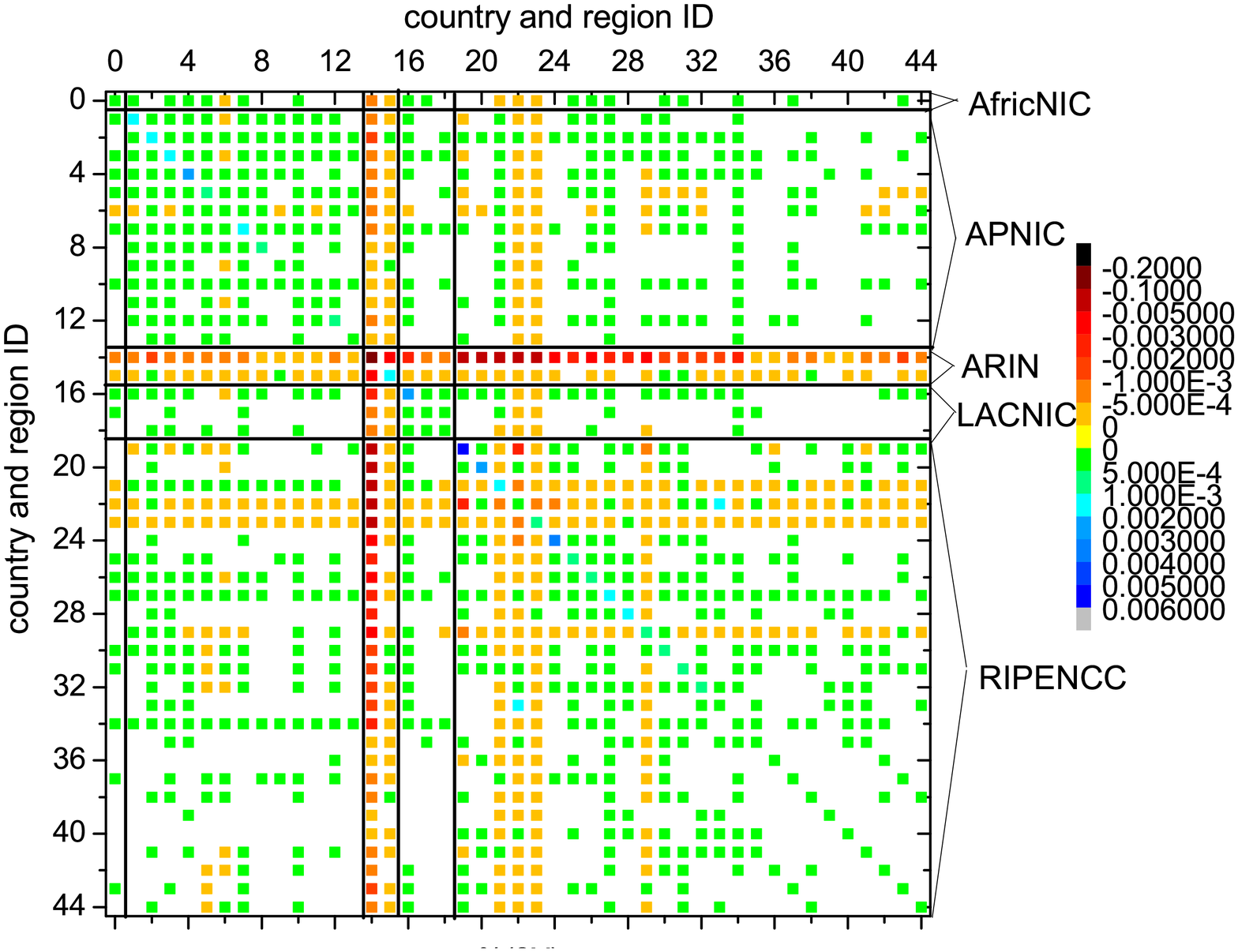}
\caption{(Color online) Connection tendencies between and within
different countries and regions.\label{country}}
\end{figure*}


\begin{table}
\centering \caption{Country and region codes for corresponding IDs
in Fig.~\ref{country}, and the number of ASes owned by these
countries and regions. }
\begin{tabular}{|c|c|c|c|}
\hline \hline
RIR name      &  ID & country and  region code &   number of ASes \\

\hline
AfriNIC          &         0  & ZA  & 95\\
\hline
\multirow{13}{*}{APNIC} &  1 & AU   & 586 \\
                        &  2  & KR  & 539 \\
                        &  3  & JP  & 483 \\
                        &  4  & ID  & 339 \\
                        &  5  & IN  & 306 \\
                        &  6  & HK  & 194 \\
                        &  7  & CN  & 166 \\
                        &  8  & TH  & 162 \\
                        &  9  & NZ  & 151 \\
                        &  10 & SG  & 123 \\
                        &  11 & PH  & 118 \\
                        &  12 & TW  & 100 \\
                        &  13 & BD  & 85 \\
\hline
\multirow{2}{*}{ARIN}   &  14 & US  & 10406 \\
                        &  15 & CA  & 674 \\
\hline
\multirow{3}{*}{LACNIC} &  16 & BR  & 574 \\
                        &  17 & AR  & 141 \\
                        &  18 & MX  & 133 \\
\hline
\multirow{26}{*}{RIPENCC}& 19  & RU & 2544 \\
                        &  20  & UA & 1146 \\
                        &  21  & GB & 1059 \\
                        &  22  & EU & 1016 \\
                        &  23  & DE & 904 \\
                        &  24  & PL & 882\\
                        &  25  & CZ & 505  \\
                        &  26  & FR & 431 \\
                        &  27  & IT & 423 \\
                        &  28  & BG & 349 \\
                        &  29  & NL & 348 \\
                        &  30  & CH & 324 \\
                        &  31  & SE & 306 \\
                        &  32  & AT & 268 \\
                        &  33  & RO & 243 \\
                        &  34  & ES & 213 \\
                        &  35  & TR & 170 \\
                        &  36  & LV & 150 \\
                        &  37  & IL & 149 \\
                        &  38  & DK & 141\\
                        &  39  & SI & 125 \\
                        &  40  & HU & 123 \\
                        &  41  & IR & 117\\
                        &  42  & FI & 111 \\
                        &  43  & BE & 110\\
                        &  44  & NO & 105 \\
\hline
\end{tabular}
\end{table}

Table.~\ref{inter-RIR} reports both the intra-RIR and inter-RIR
assortativity coefficients.  We observe that except for ARIN, other
RIRs all show assortativity internally. For inter-RIR connections,
we observe that connections between ARIN and all other RIRs show
dissortativity. RIPENCC exhibits similar phenomenon with ARIN except
that its connections with AfriNIC exhibits some sort of
assortativity. Connections among AfricNIC, APNIC, and LACNIC, all
show assortativity. This connectivity tendency reflects the fact
that broadly, the regions covered by ARIN and RIPENCC are the core
of the Internet. However, RIPENCC differs from ARIN in the sense
that RIPENCC itself is assortative whereas ARIN is dissortative.
This could be more appropriately explained by the more fine-grained
country and region connection tendencies. Fig.~\ref{country} reports
the intra- and inter-country and region assortativity coefficients
for those countries and regions whose observed ASN numbers are
greater than 80(we choose 80 as a threshold because we want to
ensure that each RIR has at least one country or region in this
map). In this figure, vacant grid means there  is no observed AS
connections between the two countries/regions. Different colors are
used to discretize the strength of assortativity/dissortativity
within and between countries/regions. Several clear patterns can be
observed from this plot. Firstly, except for US, all other
countries/regions are internally assortatively mixed, as illustrated
by the diagonal of the plot. Secondly, there are a few
countries/regions, namely, US, CA, GB, EU, DE, that primarily show
dissortative connectivity tendencies to other countries/regions.
Finally, inter-connections between other countries/regions are
mostly assortative.

Statistically, on the RIR scale, we found that about 67.3\%
intra-RIR edges are assortative, whereas only 32.3\% inter-RIR edges
are assortative. And on the country/region scale, 69.7\%
intra-country/region edges are assortative, whereas only 44.9\%
inter-country/region edges are assortative. Considering the fact
that globally an average of 60.4\% edges are assortative, it is then
apparent that on both scales, edges within the same regional area
are more likely to be assortative than the average ratio 60.4\%,
whereas, edges linking different regional areas are far less likely
to be assortative than the average ratio. This locality-driven
difference in connectivity patterns is a characteristic feature of
AS-level Internet topology, which however, cannot be revealed by the
global assortativity coefficient.

\section{Discussion and Conclusion}
Prior to our definition, local assortativity coefficient is proposed
as a local metric~\cite{piraveenan2008,piraveenan2010} that
measures the individual node's connection tendency, which is defined
by calculating the contribution of each node to the global
assortativity coefficient. However, the calculation is arguable because
there is no precise and unique way to deterministically quantify each
node $v$'s contribution to a combined term $U_q^2$ collectively calculated
from the edge set. For example, supposing the remaining degree of a node $v$ is $j$, there may be many forms of the contribution of $v$, such as $(j/\sum_{v\in V} j)*U_q^2$, $(j^2/\sum_{v\in V} j^2)*U_q^2$ and so on. None of these forms can justify itself.
This is because calculation of $U_q^2$ is a unified process,
which is closely related to the complex correlation of the network structure,
so we could not decompose this term into each node's contribution as if nodes
 were independent of each other. In contrast, our
definition is more straightforward in that it calculates each edge's
contribution to the global assortativity coefficient, rather than
each node's contribution to a term in the formula. As a result, our
definition completely avoids the bias issue in that
definition~\cite{piraveenan2010}. More importantly, from the edge
assortativity, we can define the universal assortativity coefficient
capable of network analysis.

To summarize,
we present a universal assortativity coefficient (UAC) which can be
used to calculate connection tendencies on any part of a network, such as communities, groups in multiple network scales.
Indeed, given that the target edge set is set to all edges, UAC is
exactly the global assortativity coefficient (GAC). In this sense,
GAC is a special case of UAC. Moreover, this definition is
deterministic, completely avoiding the bias issue accompanied with
the node-based local assortativity coefficient definition. UAC helps
to uncover individual, partial, and global assortativity patterns in
various networks. Applying UAC to real world networks, we find that
contrary to the popular expectation, most globally dissortative networks are
still dominated by assortative edges, though with weak strength.
This observation also motivates us to classify networks along two
dimensions into four categories, characterized by their global
assortativity coefficient and local assortativity statistics. It is
expected that this measure can be widely applied to various networks
such as popular online social networks, ubiquitous modern
communication networks and transportation networks, help people
uncover more hidden patterns in networks, and finally allow deep
understanding of network dynamics caused by the structural
difference discerned by the UAC.

\section{ACKNOWLEDGEMENT}
The authors thank Dr. Wenxu Wang for his valuable comments on this paper. This work is
partly supported by the National Natural Science Foundation of China
under Grant No. 61100178 and 61174152. 














\begin{figure*}\renewcommand{\figurename}{comments: }\notag
\centering
\includegraphics[width=18 cm]{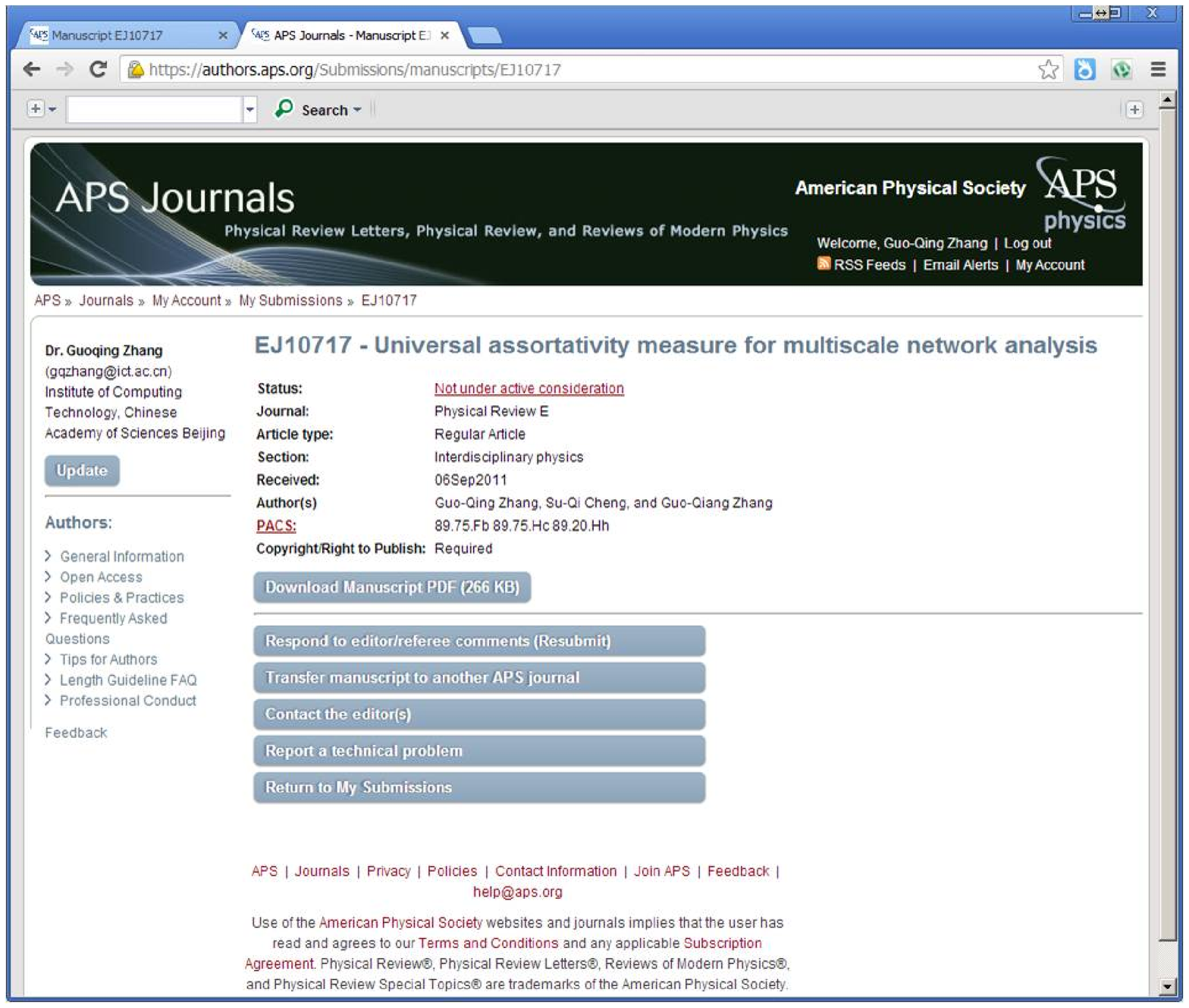}
\caption{This manuscript was submitted to Physical Review E on September 6, 2011.
}
\end{figure*}

\end{document}